\documentclass[12pt]{article}
\usepackage{times,amsmath,mathptm,citesort,epsfig}
\usepackage{graphicx}
\newcommand\bea{\begin{eqnarray}}
\newcommand\eea{\end{eqnarray}}
\newcommand\beq{\begin{equation}}
\newcommand\eeq{\end{equation}}
\begin{document}

\begin{center}
{\Large {\bf Exchange Interaction in Binuclear Complexes with Rare Earth
and Copper Ions: A Many-Body Model Study}} \\
\vspace{0.5cm}
Indranil Rudra , C. Raghu and S. Ramasesha \\
Solid State and Structural Chemistry Unit, Indian Institute of Science, \\
Bangalore 560 012, India\\
~\\
~\\
{\bf ABSTRACT}\\
\end{center}
We have used a many-body model Hamiltonian to study the nature of the magnetic 
ground state of hetero-binuclear complexes involving rare-earth and copper 
ions. We have taken into 
account all diagonal repulsions involving the rare-earth 4f and 5d orbitals and
the copper 3d orbital. Besides, we have included direct exchange 
interaction, crystal field splitting of the rare-earth atomic levels and 
spin-orbit interaction in the 4f orbitals. We have identified the inter-orbital
$4f$ repulsion, U$_{ff}$ and crystal field parameter, $\Delta_f$ as the key
parameters involved in controlling the type of exchange interaction between the
rare earth $4f$ and copper $3d$ spins. We have explored the nature of the 
ground state in the parameter space of U$_{ff}$, $\Delta_f$, spin-orbit 
interaction strength $\lambda$ and the $4f$ filling n$_f$.
We find that these systems show low-spin or high-spin ground state depending 
on the filling of the $4f$ levels of the rare-earth ion and ground state spin
is critically dependent on U$_{ff}$ and $\Delta_f$. In case of half-filling 
(Gd(III)) we find a reentrant low-spin state as U$_{ff}$ is increased, for 
small values of $\Delta_f$, which explains the recently reported apparent 
anomalous anti-ferromagnetic behaviour of Gd(III)-radical complexes. By varying 
U$_{ff}$ we also observe a switch over in the ground state spin for other 
fillings . We have introduced a spin-orbit coupling scheme which goes beyond 
L-S or j-j coupling scheme and we find that spin-orbit coupling does not 
significantly alter the basic picture.
 
\noindent

\section{Introduction}

Electron spin pairing and the associated electron exchange between atoms 
is the cornerstone of chemical bond formation. Hence most of the molecules 
have closed shell ground state \cite{hl} and the synthesis of high spin 
molecules as well as molecular magnets has been a challenge. The simplest 
idea to synthesize a high spin molecule is to start with a high spin carrier. 
In this regard lanthanides, and in particular, ${\rm Gd^{3+}}$ which has a 
$4f^7$ configuration is an obvious choice. Mononuclear complexes of 
${\rm Gd^{3+}}$ are already known. Systems with ground state spin higher 
than ${\frac{7}{2}}$ can be built up with multinuclear complexes 
involving ${\rm Gd^{3+}}$ and other rare-earth or transition metal ions. 
Studies of such heteropolymetallic compounds have also been of great 
interest in modeling metaloenzymes and in understanding their magnetic 
properties \cite{blond}.  Indeed, the interaction between two nonequivalent 
metal centers can lead to some unique features which are not encountered in 
the complexes with same kind of spin carriers. 

In the last few years, there have been several studies of heteropolymetallic
molecular coordination complexes of lanthanides and Cu(II) or other ions 
because of their luminescence and interesting magnetic properties 
\cite{kahnbook}. Some of these complexes exhibit a high spin ground state 
corresponding to a ferromagnetic alignment of magnetic moments of
the individual ions, a fact which is important for designing  building 
blocks for molecular magnets.  Gatteschi and coworkers \cite{gat1,gat2,guill} 
found that in case of Gd(III)-Cu(II) binuclear and Cu(II)-Gd(III)-Cu(II) 
trinuclear complexes, the Gd(III)-Cu(II) interaction is ferromagnetic in 
nature, independent of the structure of the complex or the ligand molecule. 
Subsequently, a systematic effort to synthesize and study binuclear systems 
between Cu(II) and other rare-earth ions was undertaken \cite{mkahn,jps1,jps2}. 
These studies indicate a general trend for the ground state spin of the 
complex as we traverse the lanthanide series. The interaction between 
Cu(II)-Ln(III) is ferromagnetic for Gd(III) and other rare-earth ions with 
greater than half-filled $4f$ shell such as ${\rm Dy(III)}$ and ${\rm 
Er(III)}$. Below half filling, this interaction is antiferromagnetic and 
leads to low-spin ground state for systems such as Cu(II)-Ce(III) and 
Cu(II)-Eu(III) \cite{mkahn}. The situation remains the same when the Cu(II) 
is replaced by organic ligands like nitronyl nitroxide triazole which are 
essentially spin-1/2 carriers. 

As a possible mechanism for the observed ferromagnetic alignment of the 
spin of the rare-earth ion with the spin of the copper ion, it was 
pointed out \cite{kahn1} that the stabilization of the high spin state arises 
from the coupling between the ground state configuration and the excited state 
configuration arising out of the transfer of the electron from the singly 
occupied copper $3d$ orbital to an empty $5d$ orbital of the rare-earth 
ion. This mechanism is in the spirit of the superexchange mechanisms
proposed by Goodenough \cite{good} in 1963. Similar mechanism is invoked 
recently by Tchougreeff \cite{tchou} to explain the ferromagnetic ordering 
in decamethylferrocenium tetracyanoethenide, by Kinoshita and coworkers 
\cite{kino} to explain the intermolecular ferromagnetic coupling in 
paranitrophenyl nitronyl nitroxide and by Girerd and coworkers 
\cite{girerd} to interpret the magnetic properties of $\mu$-oxo Mn(II) 
compounds. Indeed, the interplay of direct exchange which favours parallel
spin alignment on a given site (Hund's rule) and kinetic exchange which 
favours antiparallel alignment of spins on bonded pair of sites (by virtue of
Pauli principle) is what determines the spin state of a system \cite{bhabha}.

Kahn and coworkers modeled the superexchange process using an 
extended H\"uckel model and obtained a value for the septet-nonet 
gap from the computed J$_{{\rm Gd-Cu}}$ in fairly good agreement with 
the experimental result. But in this model strong electron-electron repulsions 
among the $4f$ electrons arising out of the contraction of the lanthanide $4f$ 
orbitals as well as between $4f$ electrons and the $5d$ electrons have not been
taken into account. These electron repulsions, according to the recent reports 
seems to play an important role in explaining the photoemission spectra of 
rare-earth compounds 
\cite{xps}. Another factor which was not accounted for in the above model is 
the splitting of the free ion Ln(III) ground state due to the ligand 
field effect, on the grounds that spin-orbit coupling term is an order of
magnitude larger than crystal field term \cite{ballh}. The need of a detailed 
microscopic model taking these factors 
into account has also been felt since some exceptions to ferromagnetic coupling 
in Gd(III) complexes were recently reported for Cu(II) complexes and nitronyl 
nitroxide radicals \cite{lescop,costes,caneschi}. To explain this unexpected
antiferromagnetic coupling, suggestions were put forward that the observed 
coupling is actually the sum of two contributions \cite{caneschi}, one from 
the direct overlap of the magnetic orbitals of the ligand with the $f$ 
orbitals, which leads to antiferromagnetism and the other from the overlap with
the $s$ and $d$ orbitals, which results in ferromagnetism. But the rare-earth 
$f$ orbitals are highly contracted to the core which prompted Kahn et. al. 
\cite{kahn1} to neglect the transfer between the $4f$ orbitals of Ln(III)
and $3d$ orbitals of Cu(II). 

In this paper we present our studies of a detailed microscopic model 
in which we have taken into account explicit electron-electron correlations, 
crystal field effects and spin-orbit effects in the Lanthanide ion. We have 
spanned the parameter space of the model and have obtained phase diagrams 
for different fillings of the $4f$ orbitals to elucidate the role played by 
the different parameters in controlling the ground state spin. The paper 
is organized as follows. In section 2, we introduce the many-body model 
Hamiltonian. We also briefly describe the computational schemes employed in 
solving the model Hamiltonian. In section 3 we discuss the results obtained 
by our computations on the model. Finally, we summarize our results in 
section 4.

\section{Many Body Model and Computational Scheme}

The remarkable feature of these systems is that the parallel or anti-parallel 
alignment of spins in the ground state is independent of the details of 
structure. This led Kahn {\it et. al.} to propose a general mechanism by the
virtue of which these systems attain either the high spin or the low spin 
state. The contraction of the $4f$ orbitals towards the nucleus means that 
they are very weakly delocalized towards the Cu $3d$ orbitals and thus, there 
is a vanishingly small overlap between the Gd $4f$ and Cu $3d$ 
orbitals. This also means that any interaction between the ground state 
configuration and the charge transfer configuration corresponding to a 
$3d-4f$ transfer cannot stabilize the low spin $S=3$ state. So, Kahn 
{\it et. al.} neglected these $3d-4f$ interactions and considered charge 
transfer to the higher $5d$ level, ie, the $3d$ electron being transfered to 
an empty $5d$ orbital of the Gd(III) ion. This process results in the $S=4$ 
high spin state being stabilized due to a strong Hund's coupling between the 
$4f$ and $5d$ levels. This process is illustrated in Fig. 1.
The mechanism outlined by Kahn {\it et. al.} is however, still in a 
non-interacting picture since it does not take into account explicit Coulomb 
repulsions among the electrons in different orbitals. Another factor missing 
in the model is the effect of the
crystal field interaction on the various orbitals involved. In order to take 
these effects into account and understand how they modify 
the ground state spin in these systems, we have studied these systems by
employing an interacting Hamiltonian. This Hamiltonian is an extension of
the Pariser-Parr-Pople model  \cite{ppp} to more than one orbital per atom
and includes direct exchange interaction terms between electrons in orbitals
on the same site.  This model has been greatly successful in describing 
ground state and low-lying excited states of various strongly correlated 
systems.

When we have more than one orbital per site, the ZDO (zero differential
overlap) approximation is not valid and we should include electron
repulsion integrals when two electrons are in different orbitals on the 
same atom. In this case, we will have several kinds of nonzero two 
electron integrals on the same site. The interaction part of the 
Hamiltonian in this modified ZDO approximation is given by,
\begin{equation}
H_{int} = 1/2 \sum_{i,j} \sum_{\mu,\mu', \nu,\nu'} [i_{\mu}i_{\mu'}|j_{\nu}
j_{\nu'}] ({ \hat E_{i \mu,i \mu'} \hat E_{j \nu, j \nu'} - \delta_{ij} 
\delta_{\mu' \nu} \hat E_{i \mu, j \nu'}}) 
\end{equation}
Here, $i,~j$ are site indices and $\mu,~\nu$ correspond to the orbital on a 
particular site. The two electron integrals $[i_{\mu}i_{\mu'}|j_{\nu}j_{\nu'}]$ 
contains contributions from various kinds of situations. The case 
$\mu~ =~ \mu^\prime~ =~ \nu~ =~ \nu^\prime$ and $i=j$ corresponds to the 
electron-repulsion integral for the
electrons in the same orbital of a single site. The most general case for the
repulsion integral on the same site is realized when there are at least four
orbitals per site and $\mu~ \ne~ \mu^\prime~ \ne ~\nu~ \ne~ \nu^\prime$ and 
$i=j$. 
For the $f$ orbitals on the same site, we have assumed that (i) 
the intra-orbital repulsion
integral $[f_\mu f_\mu|f_\mu f_\mu]$ is the same for all the seven $f$ orbitals,
independent of $\mu$, and given by $U_f$, (ii) the inter $f$ orbital repulsion 
integral $[f_\mu f_\mu|
f_\nu f_\nu]$ is also independent of the particular pair of $f$ orbitals 
$\mu,~ \nu$
and given by $U_{ff}$ and (iii) that there is also a single direct exchange 
integral $J_{ff}$ between two $f$ orbitals corresponding to the integral
$[f_\mu f_\nu|f_\mu f_\nu]$. We have further assumed that all other electron
repulsion integrals involving the $f$ orbitals are zero. The direct exchange
integral although smaller than the electron repulsion integrals in 
(i) and (ii) by about an order of magnitude, is crucial to the study of 
magnetic gaps. Indeed, the small value of the exchange integral is also
reflected in the small value of the magnetic gaps. For our
rare-earth system, in most cases, we have considered all the seven $4f$ 
orbitals and one ${\rm 5d}$ orbitals for the rare-earth metal ion and a single 
${\rm 3d}$ orbital for the Cu ion. Inclusion of $5d$ orbital introduces
corresponding electron repulsion parameters $U_{5d}$, $U_{fd}$ and $J_{fd}$.
Electron repulsion integrals of the form $[i_{\mu}i_{\mu}|j_{\nu}j_{\nu}]$ 
which in our case corresponds to repulsion integral $V_{fd}$ between an 
electron in the $4f$ 
orbital of the rare-earth and an electron in the $3d$ orbital of the Cu ion
is parametrized using Ohno interpolation scheme \cite{ohno}. All other
inter-site electron repulsion integrals are neglected. 

Using the above parameters, the model Hamiltonian is written as ,
\bea
{\hat H}~=~\sum_{f=1}^{7}~\Delta_{f}~f~{\hat n_{f}}~+~\Delta_{3d}~{\hat n_{3d}}
~+~\Delta_{5d}~{\hat n_{5d}}~~~~~~~~~~~~~~~~~~~~~~~~~~~~~~~~~~~~~~\nonumber
\\
~+~t_{3d-4f}~\sum_{f=1}^{7}~\bigl({\hat E}_{f,3d}~+~ H.~c.\bigr)
~+~t_{3d-5d}~\bigl({\hat E}_{3d,5d}~+~ H.~c.\bigr)~~~~~~~~~~~ \nonumber \\
~+~\frac {U_{f}}{2}~\sum_{f=1}^{7}~{\hat n_{f}}({\hat n_{f}}-1)
~+~ \frac {U_{3d}}{2}~{\hat n_{3d}}~({\hat n_{3d}}-1)
~+~ \frac {U_{5d}}{2}~{\hat n_{5d}}~({\hat n_{5d}}-1) \nonumber \\
~+~ U_{ff}\sum_{f>f^\prime=1}^7\hat n_f \hat n_{f^\prime}+U_{fd}\sum_{f=1}^7
\hat n_f\hat n_{5d}~~~~~~~~~~~~~~~~~~~~~~~~~~~~~~~~~~~~~~~~~~~~~~~~~ \nonumber \\
~+~V_{3d-4f}\sum_{f=1}^{7}~{\hat n_{f}}~{\hat n_{3d}}~+~V_{3d-5d}~{\hat n_{3d}}
~{\hat n_{5d}}~~~~~~~~~~~~~~~~~~~~~~~~~~~~~~~~~~~~~~~~~~ \nonumber \\
~+~J_{ff}~\sum_{f>f^{\prime}=1}^{7}~\bigl[\bigl({\hat E}_{f,f^{\prime}}~
{\hat E}_{f,f^{\prime}}~+~{\hat E}_{f,f^{\prime}}~{\hat E}_{f^{\prime},f}~+
~H.~c.\bigr)~-~ {\hat n_{f}}~-~{\hat n_{f^{\prime}}}\bigr]~~~~~ \nonumber \\
~+~J_{fd}~\sum_{f=1}^7 \bigl[\bigl({\hat E}_{f,5d}~{\hat E}_{f,5d}~+
~{\hat E}_{5d,f}~{\hat E}_{f,5d}~+~H.~c.\bigr)~-~{\hat n_{f}}~-~{\hat n_{5d}}
\bigr]~~~~~~~
\eea
\noindent
where we define the electron hop operator $\hat{E}_{ij}$ which hops an
electron of either spin from orbital$j$ to orbital $i$ as $\hat{E}_{ij} = 
\sum_{\sigma}{\hat{a}_{i,\sigma}}^{\dagger}{\hat{a}_{j,\sigma}^{}}$; $\hat 
a^\dagger_{i,\sigma}$ ($\hat a_{i,\sigma}^{}$) creates (annihilates) 
an electron 
with spin $\sigma$ in the $i^{th}$ orbital. The first two lines in Eqn. 2 
corresponds to the noninteracting part of the Hamiltonian. The first line 
corresponds to the site energy contribution. Here, we assume that the 
degeneracy of the $4f$ levels are lifted uniformly by the nonsymmetric 
crystalline field, leading to equally spaced levels with an energy level 
spacing $\Delta_f$. The $5d$ orbital of the rare-earth and the $3d$ orbital 
of the Cu are located at energies $\Delta_{5d}$ and $\Delta_{3d}$ respectively. 
The second line in the Hamiltonian corresponds to the transfer part between 
the individual $4f$ orbitals and the Cu $3d$ orbital, with a transfer 
parameter $t_{3d-4f}$ and a transfer between the rare-earth $5d$ orbital 
and the Cu $3d$ orbital with a transfer parameter $t_{3d-5d}$. Here, we 
assume that the orbitals on the same site are eigenstates of the site 
Hamiltonian and therefore do not mix. The third line in the Hamiltonian 
corresponds to Hubbard electron repulsion terms, $U_f$, $U_{5d}$ and 
$U_{3d}$ being the repulsion parameters for the orbitals we have considered. 
The fourth line involves inter-orbital interactions on the rare-earth site 
with $U_{ff}$ for repulsion between electrons in different $4f$ orbitals
and $U_{fd}$ for repulsion between electrons in the $4f$ and $5d$ orbitals
at the rare-earth site. The fifth line in the Hamiltonian corresponds
to the inter-site interaction between electrons on the rare-earth site and
those in the Cu $3d$ orbital. The parameters $V_{3d-4f}$ and $V_{3d-5d}$ are
obtained using Ohno parametrization with intersite distance of 2\AA. 
The last two lines in the Hamiltonian correspond to the exchange term 
involving orbitals on the rare-earth site. The exchange term has both diagonal
and off-diagonal contributions, the latter combing from two successive electron
hops.

We set up the matrix for the above Hamiltonian in the Valence Bond (VB) basis.
The VB basis being eigenstates of the total spin operator is ideal for this 
study since the total spin of the targeted state is known exactly. 
The VB method is discussed extensively in \cite{sr1}. The resulting 
Hamiltonian matrix is non-symmetric and is diagonalized using Rettrup's 
algorithm \cite{ret}. The low energy states are determined for a range of 
values of the parameters in the Hamiltonian. These studies were carried
out at various $4f$ orbital fillings, corresponding to different lanthanide 
ions. 

We have also carried out calculations that includes spin-orbit interactions
between the $4f$ orbital angular momentum and the electron spin. The 
spin-orbit contribution is given by,
\bea
{\hat H_{S.O.}} ~=~ \sum_{i}~\xi(r_{i})~l(i)~\cdot~s(i)
\eea
\noindent
where $\xi(r_{i})$ is given by
\bea
\xi(r_i)~=~{\frac {\hbar}{2m^2c^2r}}~{\frac {\partial V} {\partial r}}, 
\eea
with usual notations. Integrating over the radial part leads to the spin-orbit
interaction parameter $\lambda$ and ${\hat H_{S.O.}}$ can be written in
the second quantized form as,
\bea
{\hat H_{S.O.}} ~=~ \frac {\lambda}{2}~\sum_{m}~\bigl[\sqrt{l(l+1)~-~m(m+1)}~
{\hat{a}_{m+1,{\beta}}}^{\dagger}~{\hat{a}_{m,{\alpha}}}  \nonumber \\
~+~\sqrt{l(l+1)~-~m(m-1)}~{\hat{a}_{m-1,{\alpha}}}^{\dagger}~
{\hat{a}_{m,{\beta}}} ~~~~~~~~~~\nonumber \\
~+~m\bigl({\hat{a}_{m,{\alpha}}}^{\dagger}~{\hat{a}_{m,{\alpha}}}
~-~{\hat{a}_{m,{\beta}}}^{\dagger}~{\hat{a}_{m,{\beta}}}\bigr)\bigr]~~~~~~~~~~~~~~~~~~~~~~~ 
\eea
\noindent
where, $l$ takes the value 3, corresponding to the $f$ orbitals and $m$ 
varies from -3 to 3. In the presence
of spin-orbit interactions, the total spin of the system is not conserved.
Thus, we cannot employ the VB method for their solution. Instead we employ 
a constant $M_s$ basis, typified by Slater determinants, to set-up the
Hamiltonian matrix. The Hamiltonian matrix in this case is symmetric and
Davidson algorithm \cite{davidson} gives the low-lying states. Since the
states are now labelled by the total angular momentum $J=L+S$, we compute
the expectation value of $\hat S^2$ and use the nearest integer or half-integer
value $S$ such that the expectation value is nearest to $S(S+1)$. This
allows identifying the approximate total spin of the states.

\section{Results and Discussion}

The Hamiltonian in Eq. (2) has several parameters and exploring the
phase space for all possible filling of the $4f$ rare-earth orbital is 
very compute intensive. In order to meaningfully explore the parameter 
space, we have firstly identified the most sensitive parameters and 
fixed the remaining parameters at generally accepted values. We have 
considered three different models for the Ln-Cu system. These correspond 
to (a) ignoring the $5d$ levels on the rare-earth ion, (b) including all 
the five $5d$ orbitals on the rare-earth ion and (c) including only one 
$5d$ orbital on the rare-earth ion. In all these cases, we have found that
the inter-orbital electron repulsion strength , $U_{ff}$, in the $4f$ orbital
and the crystal-field splitting of the $4f$ orbitals $\Delta_f$ are the
parameters to which the spin of the ground state is most sensitive. The 
remaining parameters have been held fixed at the following values. The value of
the exchange constants $J_{ff}$ and $J_{fd}$ are fixed at 0.1$U_f$ and 0.2$U_f$
, respectively. The $U_{3d}$  on copper is
fixed at 5eV, $t_{3d-4f}$ at 0.1eV and in all the models and in models (b) and 
(c) $t_{3d-5d}$ was fixed at 0.5eV, $U_{5d}$ at 5eV and $U_{fd}$ at 4 eV. In 
what follows, we discuss our results, first for the half-filled $4f$ shell and 
then for rare-earth ions with other $4f$ occupancies, for various values of 
the parameters $U_{ff}$ and $\Delta_f$. In the last subsection we explore
the effect of spin-orbit interactions on the "spin" of the ground state for 
these parameter values.

\subsection{$4f^7$ system}

For the half-filled Ln ion in model (a) where the $5d$ orbitals on the 
rare-earth site is ignored, we always obtain a low-spin ground state, 
for all values of $U_f$ and $\Delta_f$ that we scanned. Changing the 
parameters $t_{3d-4f}$ and $U_{3d}$ within reasonable limits still 
leaves the system in a low-spin ground state. This is because, the path 
way for delocalization of the electrons in the high-spin state is blocked 
(Fig. 2a), depriving the system of kinetic stabilization. Thus, this 
model does not produce a high-spin ground state for any value of 
$\Delta_f$ or $U_{ff}$, although experimentally most $4f^7$-radical or Cu(II)
systems are found in high-spin ground state.  

In the presence of the empty $5d$ orbitals (Fig. 2b), the Hunds rule 
stabilizes the state in which the $4f$ and $5d$ electrons have the same 
spin alignment.  This in turn implies that the high-spin state in which 
the unpaired electron in the Cu $3d$ orbital has the same orientation as 
the spins in the $4f$ orbital has lower energy than the low-spin state. 
Our studies on model (b) is therefore expected to produce a high-spin 
ground state. However, what is quite unexpected is that we find a very 
interesting variation in the ground state spin; the ground state is a 
high-spin state for small $\Delta_f~ (\le 0.1eV)$ and $1.5 eV \le U_{ff} 
\le 4.5eV$ at $U_f=10eV$. For other values of $U_{ff}$, the ground state 
is in a low-spin state. Thus, in the quantum phase diagram (Fig. 3), 
there is a re-entrant low-spin phase. For small values of $U_{ff}$ the 
virtual low-spin state in which the intermediate rare-earth ion 
configuration is $4f^8$ has lower energy than the configuration $4f^75d^1$ 
and this would lead to a low-spin ground state of the complex. At 
intermediate values of $U_{ff}$, the virtual state $4f^75d^1$ has a 
lower energy than the $4f^8$ virtual state and we should expect the 
ground state to be a high spin state.  For large values of $U_{ff}$, the 
ground state configuration of the rare-earth ion is no longer $4f^7$ but 
$4f^65d^1$. In this case, the hopping between the $5d$ and $3d$ orbitals 
favours antiparallel alignment of spins. The $3d$ to $4f$ hop prefers 
parallel alignment on energy consideration but there is only one channel for 
this process as against seven channels for antiparallel alignment. 
Therefore, the ground state switches back to low-spin state for high 
values of $U_{ff}$.

We also find that the high-spin state is quite stable and continues to be 
the ground state, even when the $t_{3d-4f}$ transfer integral is increased 
to unto twice its value of $0.1eV$ we have used in all our studies. 
Furthermore, we find that the $\Delta_f$ parameter which corresponds 
to the crystal field splitting of the $4f$ levels also plays a significant 
part; when $\Delta_f$ is small, the cross over from high-spin ground state 
to low-spin ground state occurs at a higher value of $U_{ff}$. Above a 
critical value of $\Delta_f$, the system is found only in the low-spin 
ground state (Fig. 3).

This picture is not changed in any significant way, in going to model 
(c) where only one $5d$ orbital is used instead of the five $5d$ 
orbitals on the rare-earth considered in model (b). In view of this, 
in cases away from half-filled $4f$ levels, we have considered only a 
single $5d$ orbital (model (c)) for reasons of computational feasibility. 

\subsection{Systems with non-half-filled $4f$ shell}

We have studied systems with various $4f$ orbital fillings using the 
Hamiltonian in Eq. 2 and the general result we find is that when $U_{ff}$ is
small, systems with fewer than seven $4f$ electrons have a 
high-spin ground state while those with more than seven $4f$ electrons
have a low-spin ground state (Fig. 6). This result follows even in 
model (a) where we have neglected the participation of the $5d$ orbitals in
delocalization. This result is quite simply understood from the direct
exchange interaction term. In the less than half-filled case (Fig. 4), 
an electron hop from the Cu $3d$ orbital to the rare-earth $4f$ orbital 
can result in a high-spin or a low-spin rare-earth ion. If 
the unpaired electrons on the rare-earth have the same alignment as 
the hopping electron, we obtain the high-spin state, otherwise the 
resulting state of the rare-earth ion is a low-spin state. The direct 
exchange interaction within $4f$ orbitals ensures that the virtual high-spin 
state of the ion has lower energy than the virtual low-spin state. Therefore, 
the ferromagnetic alignment of the spins on the Ln-Cu system is favoured 
over antiferromagnetic alignment in the less than half-filled case. In 
the more than half-filled case, electron-hop is only permitted when the 
hopping electron has an alignment opposite that of the rare-earth spin 
(Fig. 5), because of Pauli exclusion principle. Thus, in the more than
half-filled case, we find a low-spin ground state. Indeed, this picture
is not significantly changed when we go to model (c) where we have
the copper $3d$ electron having an additional delocalization path-way
via the rare-earth $5d$ orbital. What we also note is that the $t_{3d-5d}$
does not have a significant effect, unlike in the half-filled case.

We should expect the crystal field splitting parameter, $\Delta_f$, of 
the $4f$ orbitals to play an important role in determining the ground
state of the Ln-Cu system, since in case of large $\Delta_f$, the rare
earth ion ground state would not be a high-spin ground state. For a 
similar reason, we should also expect the inter-orbital repulsion, 
$U_{ff}$ also to determine the magnetic state of the dimer. In Fig. 6,
we present some typical quantum phase diagrams for the less than
half-filled case as well as the more than half-filled case. In the
less than half-filled case, for large $U_{ff}$, we obtain a low-spin
ground state since the energy difference between high-spin and
low-spin rare-earth virtual states obtained on electron transfer,
is not significant. However, number of low-spin virtual states are
more than the high-spin virtual states and we obtain a low-spin ground 
state for large $U_{ff}$. The experimental system with less than half-filled
$4f$ shell are all known to be in the low-spin (afm) ground 
state. Thus, these systems all correspond to the large $U_{ff}$ limit. The 
critical value of $U_{ff}$ decreases with increase in $4f$ occupancy from 
1 to 7.

In Fig. 6, we show some typical quantum phase diagrams for the more
than half-filled case. Here again we note a quantum phase transition,
from the low-spin ground state to the high-spin ground state, at large 
values of $U_{ff}$. This transition is also qualitatively understood
in the following way. At large $U_{ff}$, the low-energy exchange path 
way is via the empty $5d$ orbitals. The direct exchange interaction
between the $5d$ and $4f$ orbitals on the rare-earth ensure that the
virtual states corresponding to high-spin state has a lower energy than
those corresponding to low-spin state. Thus, the dimer switches to a
high-spin ground state with increasing $U_{ff}$. All the experimental systems
known in this case correspond to high-spin ground state. Thus, we are in the
large $U_{ff}$ limit in real systems even in systems in which the rare earth
ion has more than half-filled $4f$ shell. This is also intuitively expected
since the $4f$ shell is chemically well isolated in rare earth compounds. The 
critical value of $U_{ff}$ in this case increases with increase in the $4f$ 
occupancy beyond half-filling.

\subsection{Role of spin-orbit interactions}

It is well known that the spin-orbit interactions in rare-earth ions
are important and often larger than the crystal field effects. we have 
considered only the spin orbit interaction for electrons in the $4f$ 
orbital of the rare-earth and have ignored it in the $5d$ orbital of
the rare-earth as well as in the $3d$ orbital of the copper. Our method
of calculation takes into account spin-orbit interactions at the 
same level as all other interactions in the Hamiltonian in Eq. 2, since
we set-up and solve the total Hamiltonian (Hamiltonians in Eq. 2 plus 
Eq. 3). Thus, our studies go beyond the usual L-S or j-j coupling schemes
in which either the spin-orbit interaction is treated perturbatively
within a given manifold of L and S or the electron repulsions are treated
perturbatively over the individual spin-orbit coupled states. We have used 
values of $\lambda$ in the range 0 to 1.0 eV, for the $U_{ff}$ and 
$\Delta_f$ values \cite{mkahn}. In the presence of spin-orbit interactions, 
total spin is no longer a good quantum number. Thus we cannot strictly 
classify the eigenstates by their spins as low-spin or high-spin states. 
However, we find that the expectation value of $\hat S^2$ in the low-lying
states, in all the cases we have studied, is close enough to allowed $S$ 
values in the absence of spin-orbit interactions. This aids in labelling the 
states as `high-spin' or `low-spin' states.

What we find is that the role of the spin-orbit interaction in determining
the ground state magnetic moment of the Ln-Cu dimers is negligible. In 
Fig. 7, we show the quantum phase diagram for three representative cases
corresponding to less than half-filled, half-filled and more than half-filled
$4f$ shells, for $\Delta_f=0$ as a function of $\lambda$ and $U_{ff}$, with
all other parameters being the same as in Figs. 3 and 6. We find that the
$U_{ff}$ value for which the magnetic transition occurs in the ground state
is hardly affected by spin-orbit interactions. Only the re-entrant low-spin
state occurs for a slightly smaller value of $U_{ff}$ when spin-orbit
interaction is sufficiently strong. We have also explored the role of 
spin-orbit interaction when the crystal field splitting of the $4f$ orbitals
$\Delta_f$ is varied near a critical value of $U_{ff}$. Here we assume that
the splitting of the $f$-orbital manifold by the crystal field does not
quench the orbital angular momentum of the resulting state. In Fig. 8 is shown 
the phase diagram as a function of $\lambda$ for the case of the rare-earth
electron configuration $4f^{10}$. We have chosen this configuration as we found
the strongest effect due to spin-orbit interactions in this case. We find 
in Fig. 7 that high-spin state crosses over to the low-spin state at a smaller
value of $\Delta_f$ when the spin-orbit interaction strength $\lambda$ is
increased. Thus, we conclude from our studies that the magnetic state
of the cluster is only weakly influenced by spin-orbit interactions.

\section{Summary and Conclusions}

In this paper, we have studied the magnetic ground state of the hetero 
bi-metallic organic molecular magnets, with a rare-earth ion and a copper 
ion. These systems have been modeled with a many-body Hamiltonian, which
includes the direct exchange interaction and the repulsion between the 
orbitals on the same site. The systems are shown to exhibit a low-spin or 
a high-spin ground state depending on the filling of electrons in the rare 
earth $4f$ levels, and also critically on the inter-orbital $4f$ electron 
repulsions, $U_{ff}$ and the splitting of the degenerate $4f$ orbitals in 
the presence of the crystal field, $\Delta_f$. We have obtained a "phase 
diagram" in the space of these parameters. We find at half-filling a 
re-entrant low-spin state as $U_{ff}$ is increased for small values of
$\Delta_f$. There is also a transition in the ground state spin at other
fillings, again for small values of $\Delta_f$, when $U_{ff}$ is increased.
The general phase diagram of the ground state spin is shown in Fig. 9, as 
a function of filling and $U_{ff}$. This picture is not altered by the 
inclusion of spin-orbit interactions. To obtain the experimentally observed
interaction, corresponding to low-spin ground state below half-filling and 
high-spin ground state above half-filling of the $4f$ shell, we need to assume
fairly large inter-orbital repulsions in the $4f$ shell. Our studies show that 
in systems where the parameters are near the phase boundaries, small changes 
brought about by the modification of the ligands can lead to drastic changes in
the magnetic properties of the complex.

\begin{center}
{\bf Acknowledgement}\\
\end{center}
We thank Prof. Diptiman Sen for many useful discussions .
We thank the Council of Scientific and Industrial Research, India and
Board of Research in Nuclear Sciences (BRNS) for financial support.

\pagebreak

\begin{figure}
\centerline{\includegraphics[height=10cm]{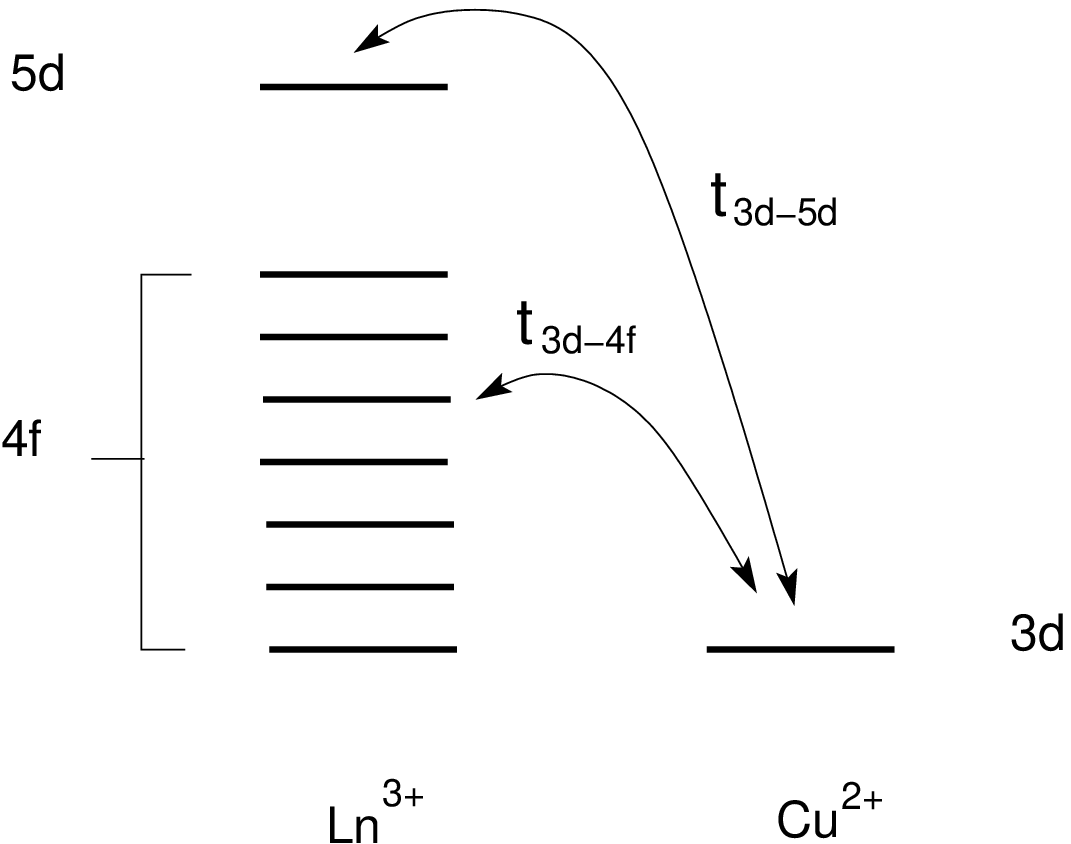}}
\vspace{1.0cm}
\begin{center}
\caption{A schematic for the mechanism which leads to the stabilization of 
the high spin state in the system. ${\rm Ln^{3+}}$ is the rare-earth 
(Lanthanide) ion. Transfers between the Cu$^{2+}$ $3d$ and the Ln $4f$ and $5d$ 
levels are indicated by the arrows. The 3d$\rightarrow$4f transfer leads to a
higher energy virtual state than 3d$\rightarrow$5d transfer.} 
\end{center}
\end{figure}

\pagebreak

\begin{figure}
\centerline{\includegraphics[height=6cm]{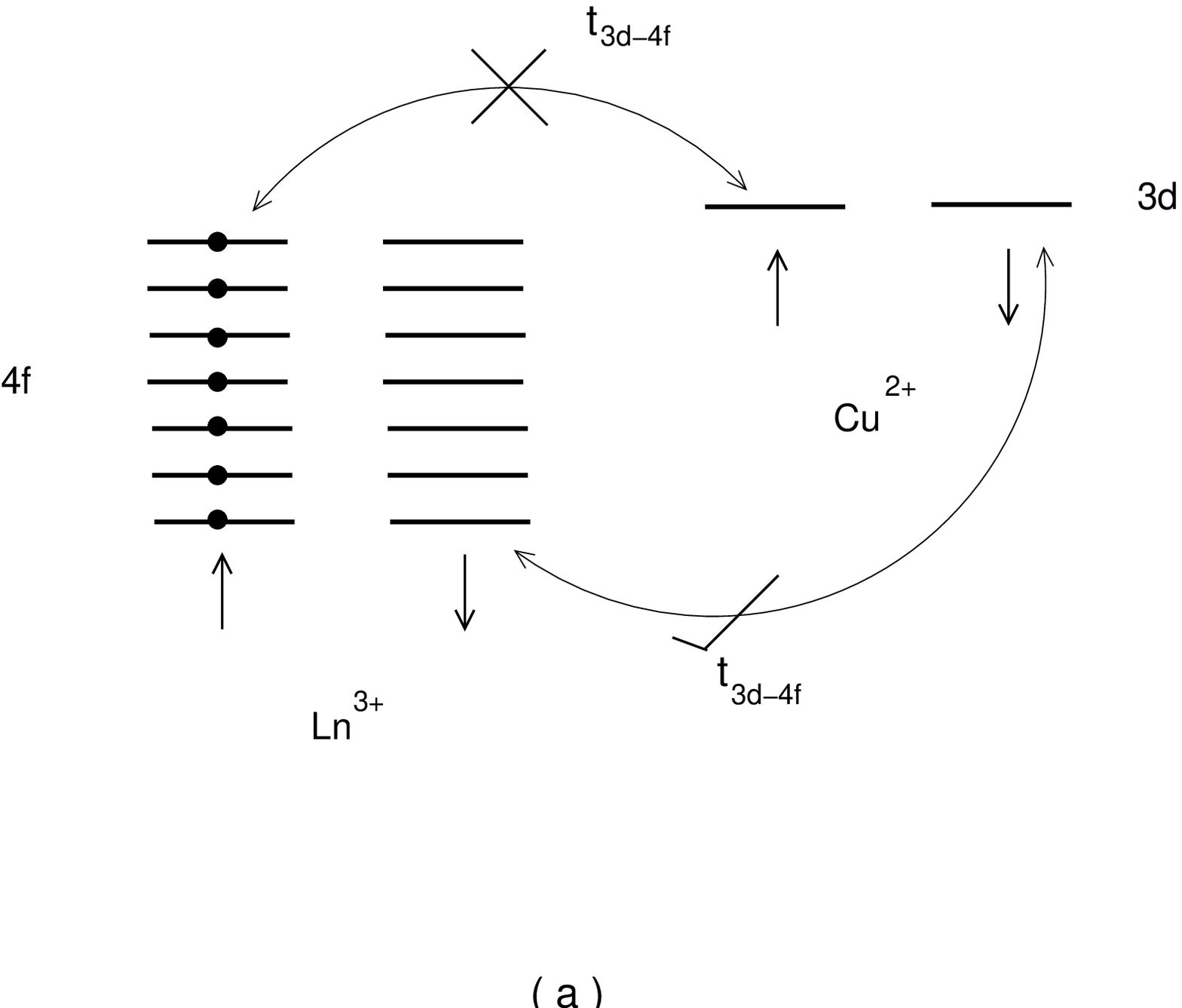}}
\vspace{1.0cm}
\centerline{\includegraphics[height=9cm]{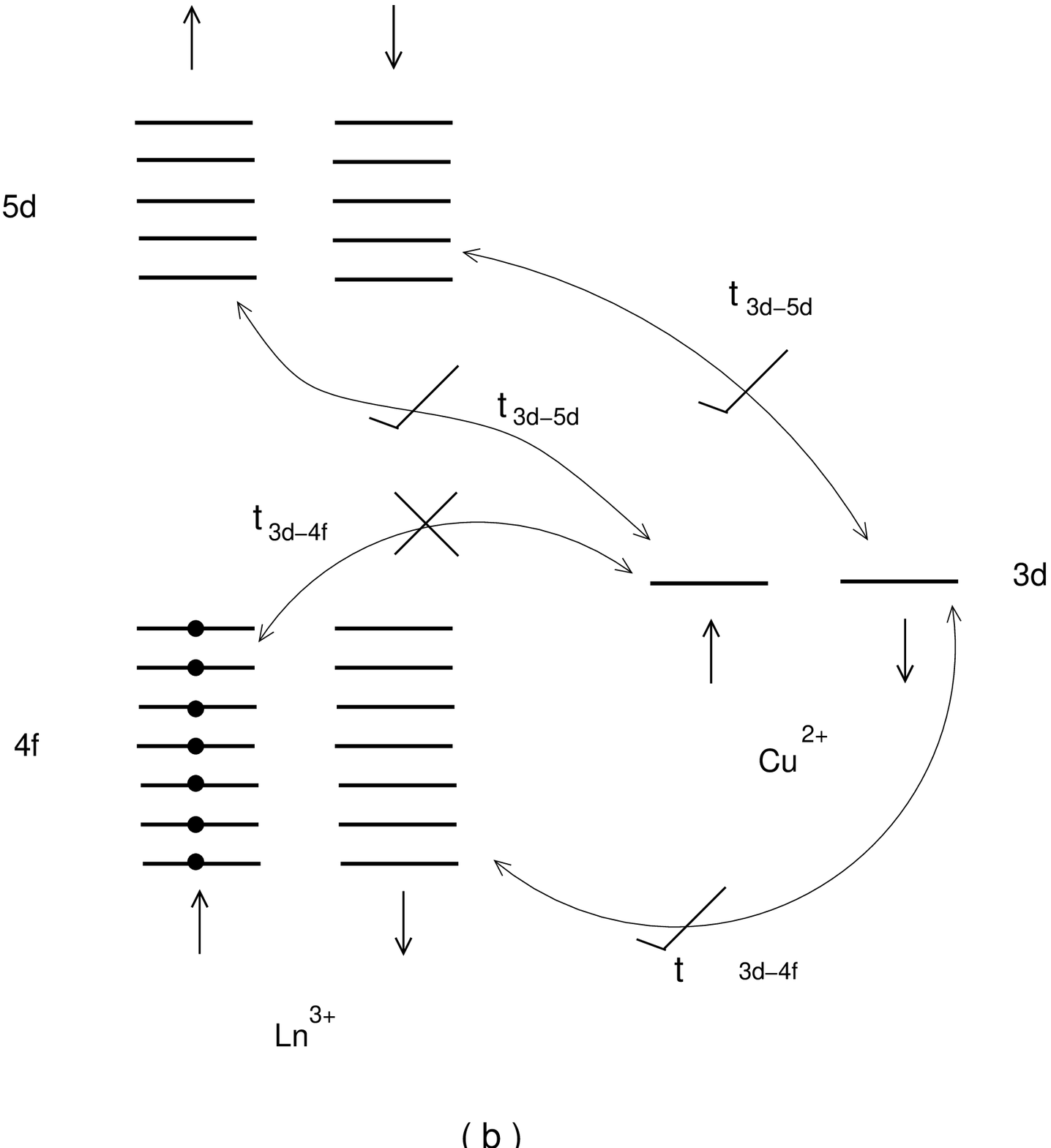}}
\vspace{1.0cm}
\begin{center}
\caption{A schematic diagram showing the various possible virtual states 
obtained by electron transfer from $3d$ orbital of Cu(II) to  orbitals of 
Ln(III), when $4f$ is half-filled - (a) without $5d$ levels, (b) with all the 
$5d$ levels taken into consideration.}
\end{center}
\end{figure}

\pagebreak

\begin{figure}
\centerline{\includegraphics[height=10cm]{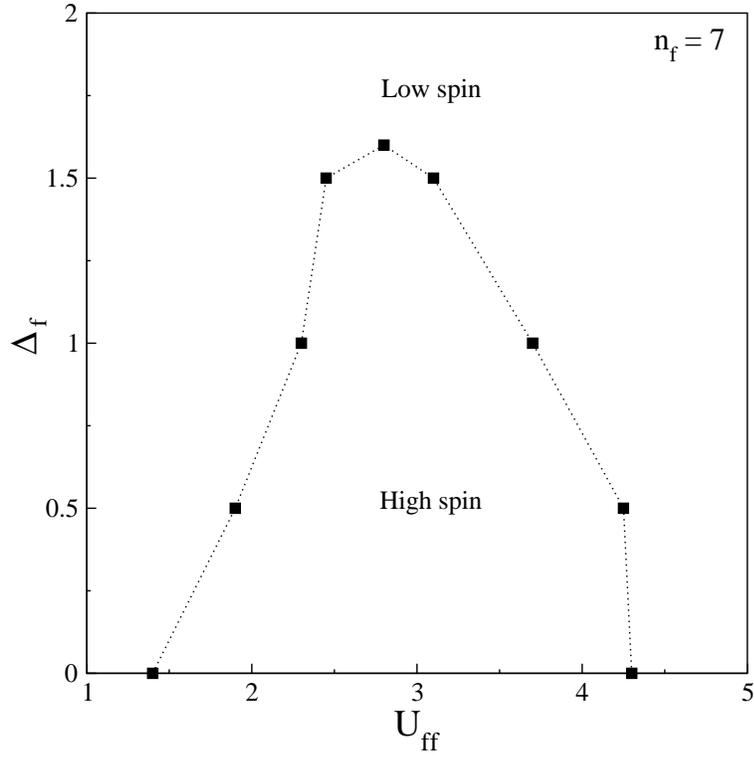}}
\begin{center}
\caption{Phase diagram in the space of parameters $U_{ff}$ and $\Delta_{f}$,
for the case of rare-earth $4f$ levels being half-filled. Other parameter
values are as follows: $U_f$=10 ev, $t_{3d-4f}$=0.1 ev, $t_{3d-5d}$=0.5 ev,
$U_{5d}$=5 ev, $U_{3d}$=5 ev, $U_{fd}$=4 ev, $J_{ff}$=0.1 ev and $J_{fd}$=0.2 
ev } 
\end{center}
\end{figure}

\pagebreak

\begin{figure}
\centerline{\includegraphics[height=10cm]{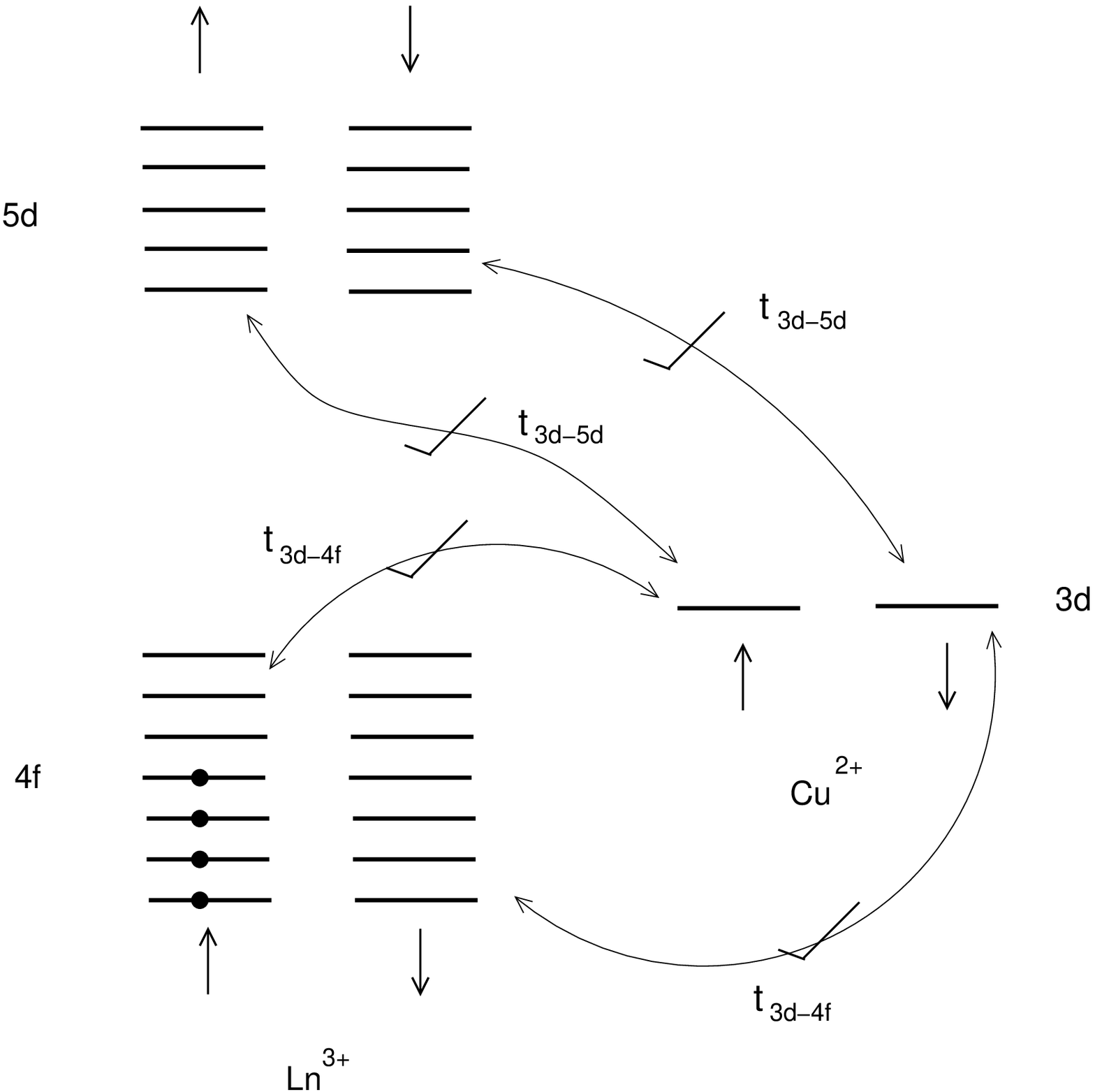}}
\begin{center}
\caption{A schematic diagram showing the various possible virtual states 
obtained by electron transfer from $3d$ orbital of Cu(II) to  orbitals of 
Ln(III), when $4f$ is less than half-filled.} 
\end{center}
\end{figure}

\pagebreak

\begin{figure}
\centerline{\includegraphics[height=10cm]{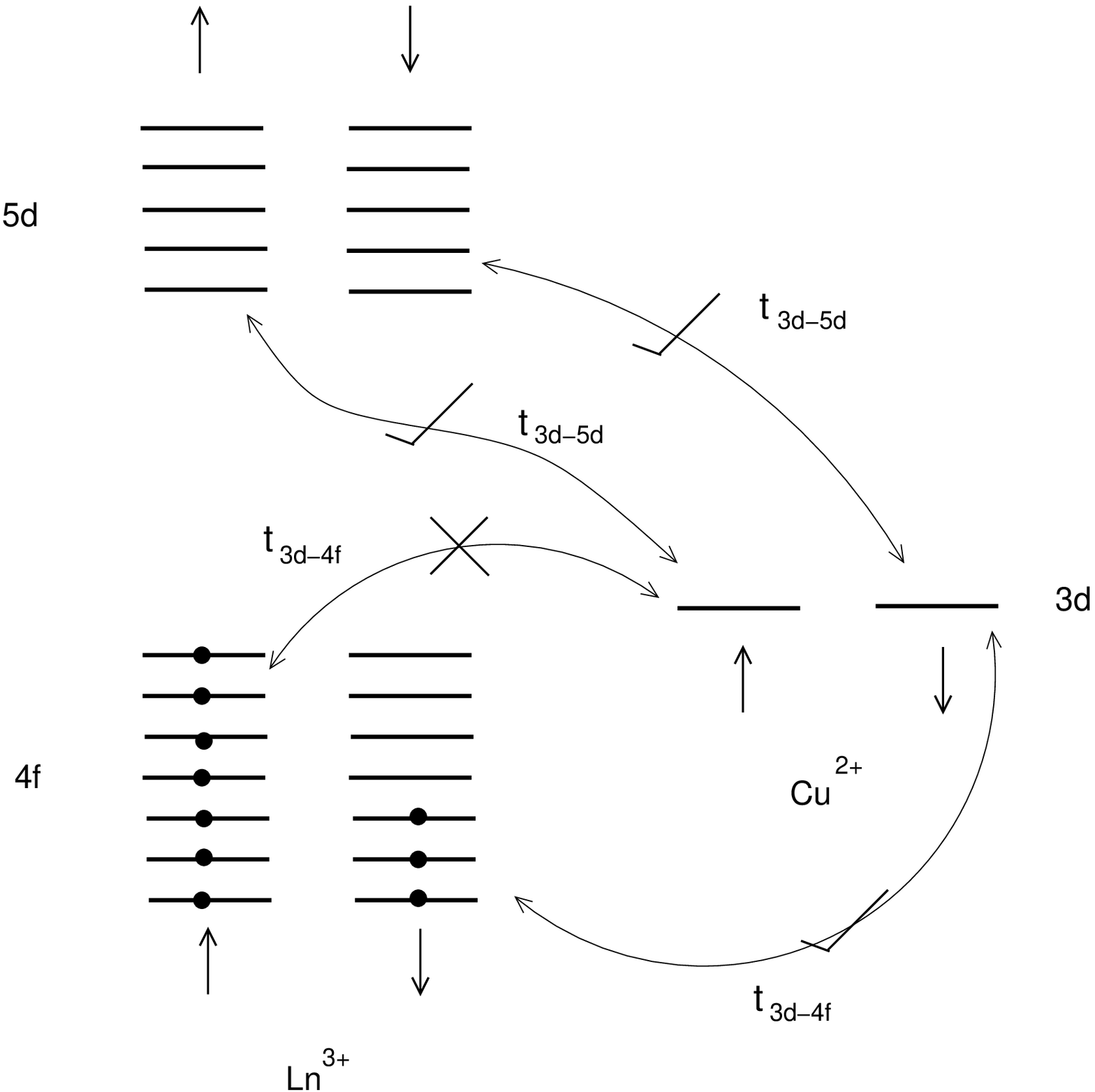}}
\begin{center}
\caption{A schematic diagram showing the various possible virtual states 
obtained by electron transfer from $3d$ orbital of Cu(II) to  orbitals of 
Ln(III), when $4f$ is more than half-filled.} 
\end{center}
\end{figure}

\pagebreak

\begin{figure}
\vspace{-3.0cm}
\begin{center}
\includegraphics[height=10cm,bbllx=26,bblly=218,bburx=515,bbury=632]{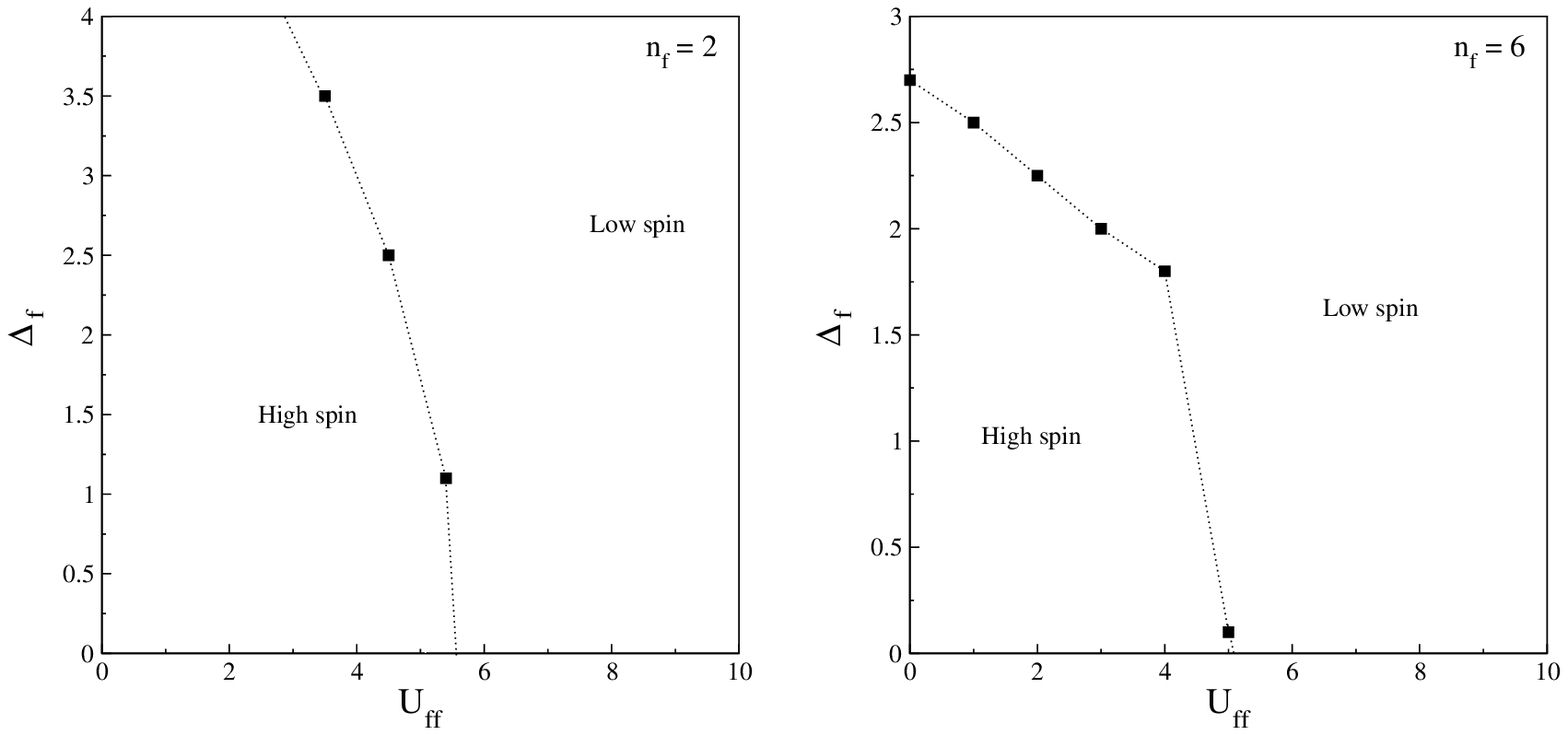}
\includegraphics[height=10cm,bbllx=26,bblly=158,bburx=515,bbury=552]{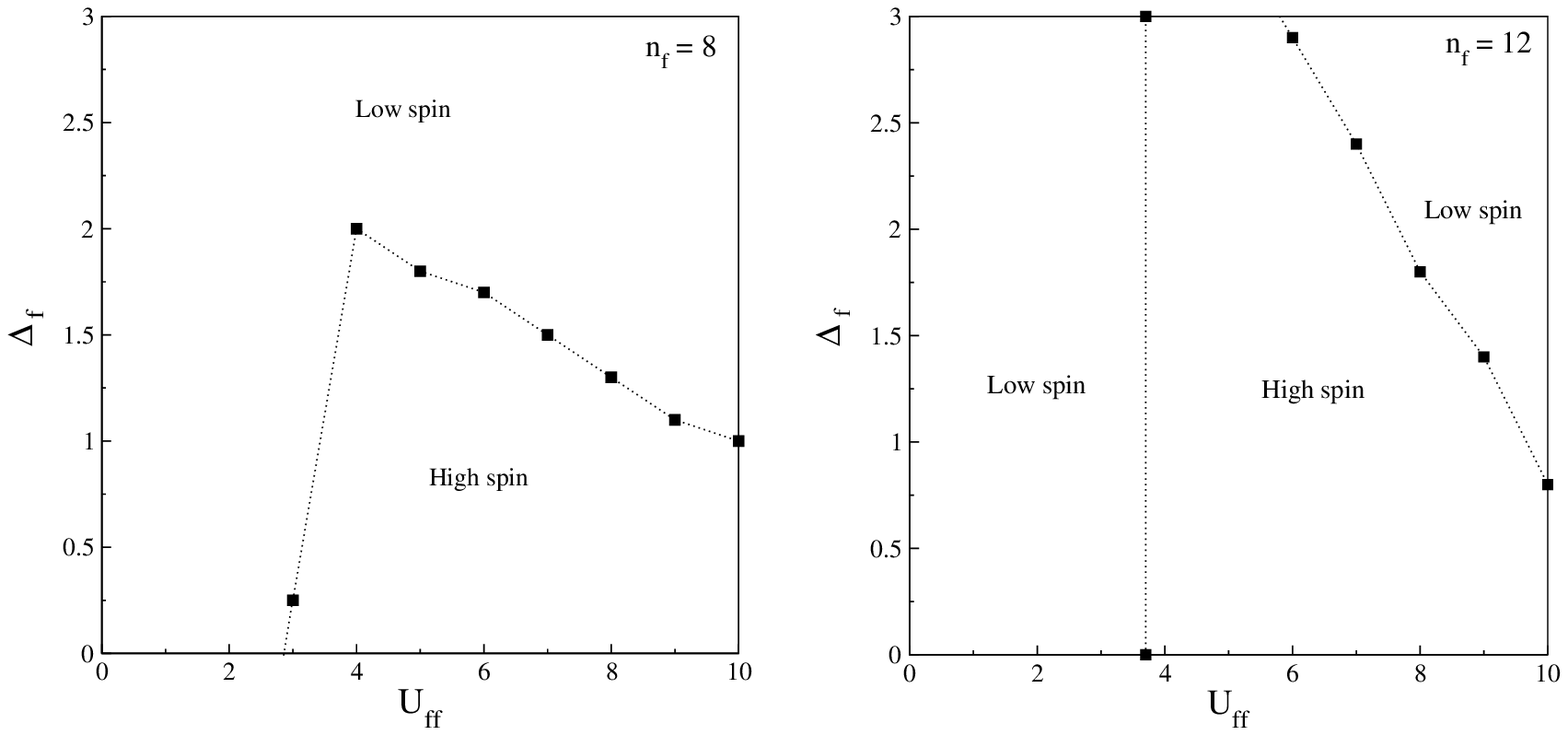}
\end{center}
\vspace{-3.0cm}
\begin{center}
\caption{Phase diagram in the space of parameters $U_{ff}$ and $\Delta_{f}$,
for the case of rare-earth $4f$ levels being away from half-filling. Other
parameter values are same as in Fig. 3}
\end{center}
\end{figure}

\pagebreak

\begin{figure}
\vspace{-3.0cm}
\centerline{\includegraphics[height=10cm,bbllx=26,bblly=158,bburx=515,
 bbury=632]{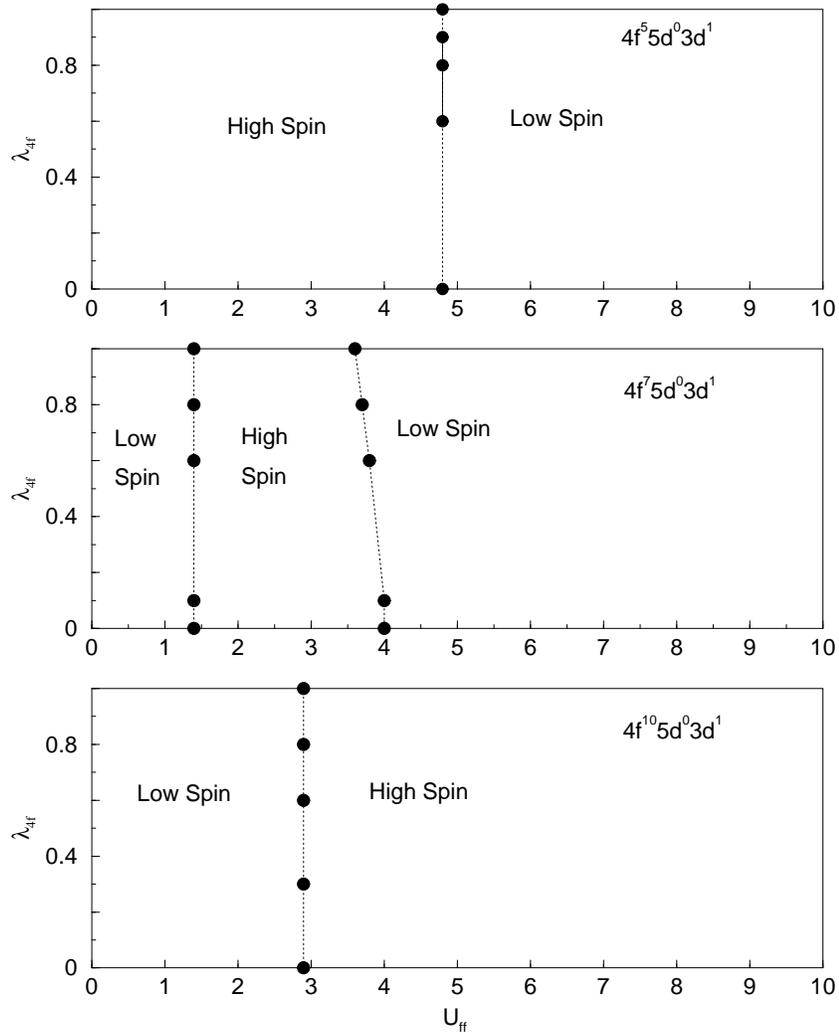}}
\vspace{3.0cm}
\begin{center}
\caption{Phase diagram in the space of parameters $U_{ff}$ and $\lambda_{4f}$,
for the case of rare-earth $4f$ levels at less than half-filling, half-filling
and more than half-filling. Other parameter values are same as in Fig. 3}
\end{center}
\end{figure}

\pagebreak

\begin{figure}
\vspace{-3.0cm}
\centerline{\includegraphics[height=10cm,bbllx=26,bblly=158,bburx=515,
 bbury=632]{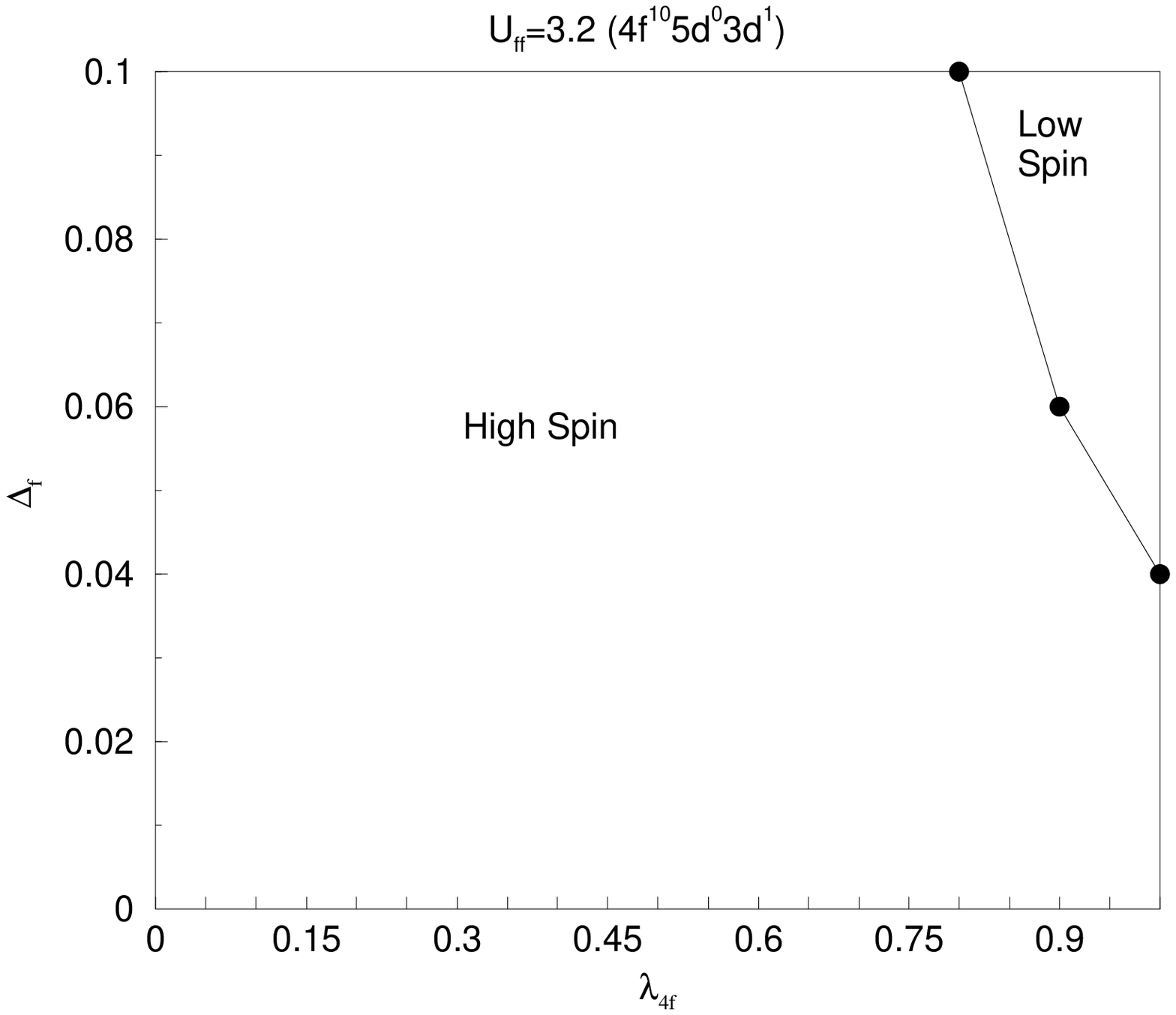}}
\vspace{3.0cm}
\begin{center}
\caption{Phase diagram in the space of parameters $\lambda_{4f}$ and 
$\Delta_{f}$ for the case of rare-earth $4f$ levels being away from 
half-filling. Other parameter values are same as in Fig. 3}
\end{center}
\end{figure}

\pagebreak

\begin{figure}
\centerline{\includegraphics[height=10cm]{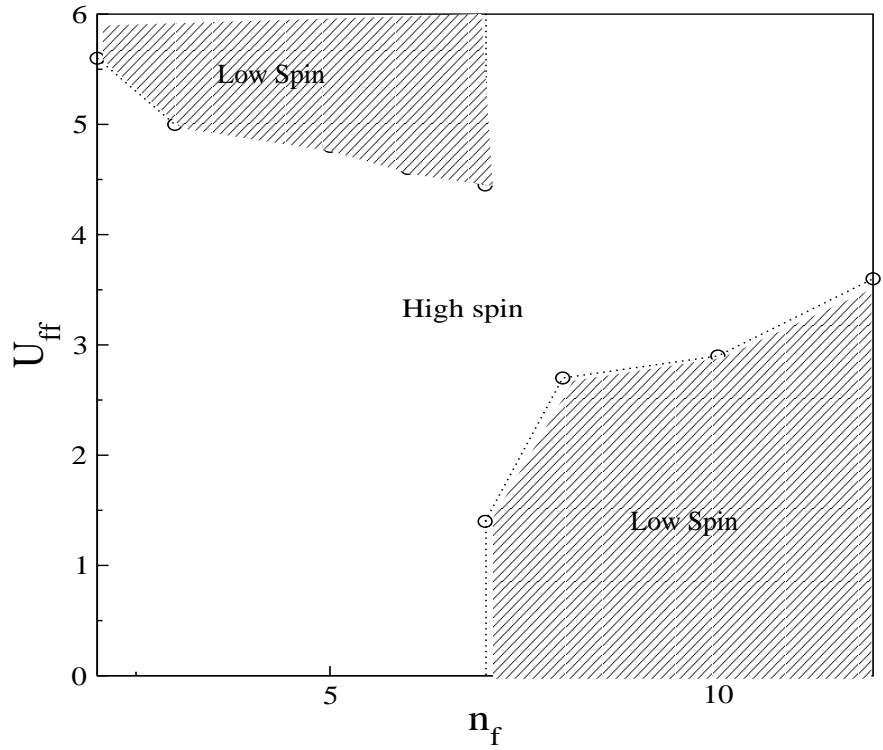}}
\vspace{1.0cm}
\begin{center}
\caption{A phase diagram for the Ln(III)-Cu(II) system, between parameters
$U_{ff}$ and $n_{f}$, $n_f$ being the $4f$ level filling. $\Delta_f$ value was
fixed at 0.1 ev. Other parameter values are same as in Fig. 3}
\end{center}
\end{figure}

\end{document}